\documentclass{article}
\usepackage{iclr2027_conference,times}
\usepackage[T1]{fontenc}
\usepackage[utf8]{inputenc}
\usepackage{amsmath,amssymb,graphicx,booktabs,tabularx,array,microtype,etoolbox}
\usepackage{hyperref}
\usepackage{url}
\hypersetup{colorlinks=true,citecolor=blue,linkcolor=blue,urlcolor=blue,pdfauthor={},pdftitle={GCAM-Gym: A Benchmark for Agent-Based Climate Policy Search}}
\title{GCAM-Gym: A Benchmark for\\Agent-Based Climate Policy Search}

\newcommand{\AuthorSep}{\hspace{1.5em}}
\newcommand{\AuthorVSep}{\\[1.0ex]}

\newcommand{\Mila}{\textsuperscript{1}}
\newcommand{\UToronto}{\textsuperscript{2}}
\newcommand{\McGill}{\textsuperscript{3}}
\newcommand{\UMontreal}{\textsuperscript{4}}

\author{
    \parbox{\textwidth}{
        \vspace{3.5ex}
        \centering
        Tianyu Zhang\Mila\AuthorSep
        Zhenghan Tai\UToronto\AuthorSep
        Xinyu Wang\McGill\AuthorSep
        Irina Rish\Mila\AuthorSep
        Yoshua Bengio\UMontreal
        \\[2.5ex]
        {
            \small\normalfont
            \Mila{}\,Mila -- Quebec Artificial Intelligence Institute\AuthorVSep
            \UToronto{}\,University of Toronto\AuthorSep
            \McGill{}\,McGill University\AuthorVSep
            \UMontreal{}\,Universit\'e de Montr\'eal
        }
    }
}

\iclrfinalcopy
\newcommand{\degreeC}{\ensuremath{^{\circ}\mathrm{C}}}
\begin{document}
\maketitle
\begin{abstract}

Comparing climate-policy search methods requires accounting for differences in policy control, available observations, feedback timing, and simulator budgets. We introduce GCAM-Gym, a Gymnasium-compatible environment that supports full-policy evaluation and stepwise control through a common interface and exposes the components of the policy objective. We compare seven search and control systems in three settings: target attainment and cost, cost-temperature trade-offs, and one-step delayed feedback. Each system is evaluated with ten seeds per setting, for a total of 3,210 attempted GCAM evaluations. Four systems share a budget of 24 attempts per seed; among these, CMA-ES achieves the highest interquartile mean (IQM) of the best returns from completed policy evaluations in all three settings. We release the evaluation archive and analysis tools to reconstruct the reported comparisons without running GCAM.

\end{abstract}

\begin{figure}[!h]
\centering
\includegraphics[width=\linewidth]{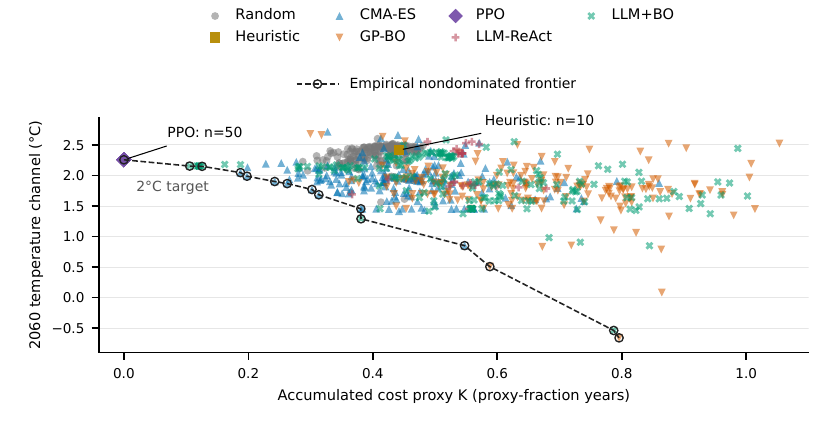}
\caption{\textbf{Different policy-search methods explore different cost--temperature outcomes.} Each point is a completed policy evaluation in GCAM-Gym, colored by method. The horizontal axis shows an accumulated policy-cost proxy; the vertical axis shows the simulator's temperature readout for 2060. Lower cost and temperature lie toward the bottom left; the horizontal dashed line marks the study's $2\degreeC$ target. Connected open circles mark outcomes for which no other recorded point has both no greater cost and no higher temperature, with at least one strictly lower. Section~\ref{sec:integrity} examines temperature-readout defects that affect these plotted outcomes. The 834 completed evaluations use method-specific budgets (Table~\ref{tab:config}); repeated outcomes overlap.}
\label{fig:tradeoffs}
\end{figure}

\section{Introduction}

Climate-policy design involves decisions whose consequences unfold over decades and interact across sectors. A carbon-price trajectory changes incentives in energy and land systems, with climate consequences mediated by emissions and the carbon cycle. Integrated assessment models provide a computational framework for studying these interactions. GCAM couples energy, water, land, climate, and economic systems~\citep{gcam}.

Searching for policies with a simulator can follow different interaction patterns. A schedule optimizer proposes a complete policy for evaluation, while a stepwise controller chooses actions as observations arrive. Comparing their performance requires specifying what each method controls, which observations it receives and when, and how many evaluation attempts it uses, including failed attempts. A benchmark for GCAM-based policy search therefore needs to support both interaction patterns and link proposed policies to executed trajectories, feedback histories, and evaluation outcomes.

We introduce \textbf{GCAM-Gym}, a Gymnasium-compatible benchmark environment for climate-policy search. Methods specify a global carbon-price trajectory, either as a complete policy or through stepwise actions, using a shared action decoder and GCAM backend. The environment returns structured observations of climate and energy-system responses, together with cost and distributional proxies. A feedback wrapper supports a one-step delay in the observations and rewards delivered to controllers, and the evaluation protocol counts both completed and failed attempts toward each system's budget. Individual reward components are returned alongside the scalar objective so that users can inspect their contributions to a policy's score. The accompanying archive links proposed policies, executed trajectories, feedback, and execution status; analysis tools reconstruct the reported comparisons without running GCAM.

We use GCAM-Gym to compare seven systems spanning fixed policies, schedule search, and observation-driven control: random search, fixed heuristics, CMA-ES, Gaussian-process Bayesian optimization, PPO pretrained on a mock backend, LLM-ReAct, and LLM-initialized Bayesian optimization. We examine whether these systems find target-reaching policies and at what cost, which cost--temperature trade-offs they attain, and how they perform under one-step delayed feedback. Each system is evaluated with ten seeds in each setting, using the configurations and budgets specified in Section~\ref{sec:protocol}.

Among the four systems evaluated with a common budget of 24 attempts per seed, CMA-ES achieves the highest interquartile mean of the best returns from completed policy evaluations in all three settings.

\section{Related Work}

\paragraph{Climate-policy environments.}
Integrated assessment models connect economic decisions to climate outcomes~\citep{dice,scc}, providing one setting for machine learning in climate mitigation~\citep{climateml}. GCAM provides the coupled simulator underlying this resource~\citep{gcam}. RICE-N studies climate negotiations, agreements, and cooperation among regional decision makers~\citep{rice}. GCAM-Gym focuses on single-agent global carbon-price control. Its evaluation unit is a policy trajectory produced through a shared simulator interface, with time-indexed responses retained for inspection. Shared Socioeconomic Pathways describe alternative futures and their energy, land-use, and emissions implications~\citep{ssp,sspscenarios}; robust decision making examines policies across uncertain futures~\citep{robust}. These differ from our three evaluation settings. The AI Economist studies taxation with coadapting agents and a planner~\citep{aieconomist}, rather than single-agent carbon-price search.

\paragraph{Reusable optimization evaluation.}
Gym standardizes environment interfaces, while the DeepMind Control Suite and bsuite provide continuous-control and diagnostic tasks~\citep{gym,dmcontrol,bsuite}. COCO and the Black-Box Optimization Challenge provide complementary optimizer-evaluation settings~\citep{coco,bobenchmark}. HPOBench demonstrates how common interfaces and isolated configurations support reproducible optimizer comparisons~\citep{hpobench}. GCAM-Gym applies this resource-oriented approach to policy trajectories in a coupled climate simulator. In addition to an objective value, an attempted evaluation produces a sequence of actions, observations, reward components, and execution statuses. This record supports both endpoint comparisons and inspection of when feedback reaches a controller.

\paragraph{Search objectives and feasibility.}
Random search provides a nonadaptive reference~\citep{randomsearch}; surrogate-based global optimization and GP-based tuning motivate sample-efficient search~\citep{ego,practicalbo}. TuRBO uses local models, and SCBO extends trust-region search to constraints~\citep{turbo,scbo}. Explicit constraint modeling and safe exploration differ from assigning penalty scores to failures~\citep{constrainedbo,safeopt}; our adapters do not inherit their feasibility or safety guarantees. Displaying attained cost--temperature pairs likewise differs from running a Pareto-search algorithm such as NSGA-II~\citep{nsga2}.

\paragraph{Language models and delayed interaction.}
Chain-of-thought prompting studies intermediate reasoning, while Reflexion retains verbal feedback across trials~\citep{cot,reflexion}. OPRO generates candidates from evaluated solutions, and LLAMBO integrates language models into BO, including warm-starting~\citep{opro,llambo}. Our LLM+BO uses language-generated initialization followed by conventional GP search, rather than reproducing LLAMBO. Research on randomly delayed interaction explicitly models action and observation delays~\citep{delay}. We instead evaluate the configured systems under a fixed one-step feedback lag.

\paragraph{Baseline methods and statistics.}
The suite uses established approaches: CMA-ES~\citep{cma}, Gaussian-process optimization through BoTorch~\citep{botorch}, PPO~\citep{ppo}, and an adaptation of the ReAct interaction pattern~\citep{react}. Their actual configurations are part of the resource, including initialization, pretraining, and failure feedback. For aggregate reporting, we use IQMs and seed-bootstrap intervals motivated by robust empirical evaluation~\citep{rliable}; the independent-unit and aggregation conventions of this study are specified below. Prior work highlights implementation and run-level variability~\citep{rlmatters,implementation}; bootstrap resampling provides the underlying statistical foundation~\citep{bootstrap}.

\section{The GCAM-Gym Interface}
\label{sec:resource}

\begin{figure}[t]
\centering
\includegraphics[width=\linewidth]{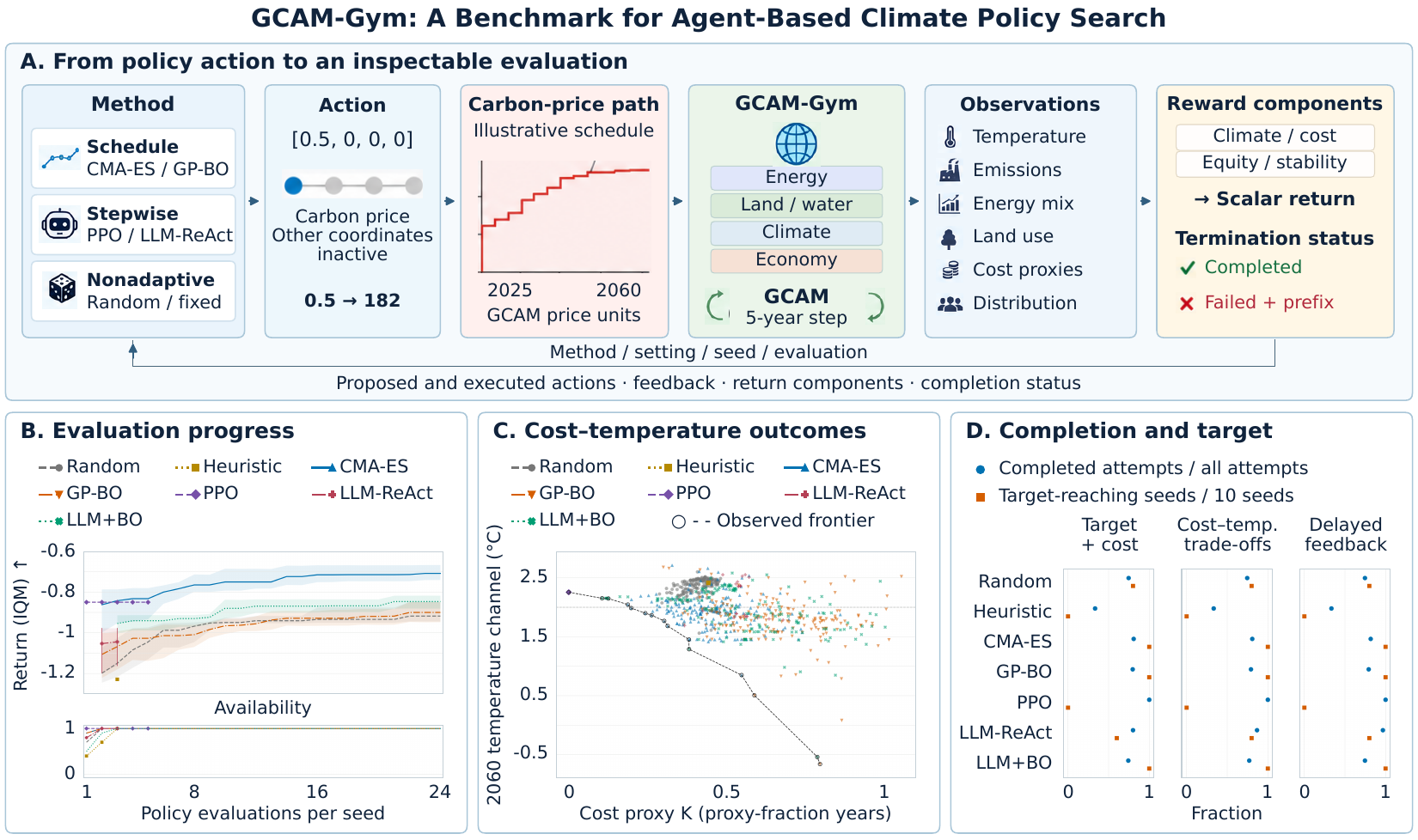}
\caption{\textbf{GCAM-Gym links policy actions to inspectable outcomes.} \textbf{A} Shared evaluation interface; the price path is illustrative. \textbf{B} Target-and-cost progress: best-return IQM with pointwise 95\% seed-bootstrap intervals, and the fraction of ten seeds with a completed policy. Failed evaluations count toward the budget; PPO is evaluated, not trained. \textbf{C} All 834 completed cost--temperature evaluations; lower-left means lower cost and temperature. The horizontal dashed line marks $2\degreeC$; connected circles show nondominated outcomes within the recorded sample. \textbf{D} Completion and target attainment across three settings. Budgets differ (Table~\ref{tab:config}); total computation is not matched.}
\label{fig:overview}
\label{fig:interface}
\end{figure}

\subsection{State, observation, and action spaces}

The environment implements the Gymnasium \texttt{reset}/\texttt{step} interface~\citep{gymnasium}. The backend initializes the simulator and executes each period; the environment decodes actions, assembles observations, and returns rewards and episode-end flags (Figure~\ref{fig:interface}A). We evaluate policies with GCAM 9.1 from an initial state in 2025 to an endpoint in 2060, using $H$ decision periods of $\Delta t$ years each. Table~\ref{tab:interface-parameters} lists the configuration. A stylized mock backend supports interface development and PPO pretraining; all reported policy evaluations use the native GCAM backend.

\paragraph{State and observation spaces.}
The internal simulator state $x_t\in\mathcal{X}$ comprises the coupled model variables and execution state. Agents receive an observation $o_t\in\mathcal{O}$: a dictionary of twelve climate, energy, cost, and distributional fields that flattens to 22 scalars (Table~\ref{tab:schema}). These observations summarize the simulator response; their Markov sufficiency is not assumed. The solver-iteration field is a constant-zero placeholder, and execution status is returned separately. Controllers can use the dictionary, a flattened vector, or a fixed-key-order serialization for an LLM. Appendix~\ref{app:interface} gives the declared bounds and field ordering.

\paragraph{Action spaces.}
The API action space is $\mathcal{A}_{\mathrm{API}}=[0,1]^4$, with coordinates for carbon price, low-carbon subsidy, coal phaseout, and carbon removal. Only carbon price affects the native simulator in this study, giving an effective normalized control space $\mathcal{A}_{\mathrm{native}}=[0,1]$. Nonzero inactive coordinates, which can occur in PPO outputs, are ignored by the native backend. The active coordinate $a_t$ maps to an integer price $p_t\in\{0,\ldots,p_{\max}\}$ through
\begin{equation}
 p_t=\operatorname{round}\left[
 p_{\max}\frac{\exp(\beta a_t)-1}{\exp(\beta)-1}\right].
 \label{eq:action-decoder}
\end{equation}
Here $p_{\max}$ sets the price ceiling and $\beta$ controls the curvature of the map. The backend treats $p_t$ as a price per tonne of CO$_2$ and writes $\kappa p_t$ to the native carbon market, using the carbon-mass conversion factor $\kappa$ in Table~\ref{tab:interface-parameters}. The native market is documented in 1990 dollars per tonne of carbon. The interface performs no currency-year adjustment, and the monetary units of the input datasets have not been independently validated (Appendix~\ref{app:interface}).

A schedule optimizer selects $(a_1,\ldots,a_H)\in[0,1]^H$ before execution; a stepwise controller selects each action from the available feedback. Both use the same decoder and period execution. Search-domain restrictions are system-specific: random search and LLM initialization use monotone schedules, while CMA-ES and subsequent BO proposals use the full normalized schedule domain (Appendix~\ref{app:methods}).

\paragraph{Interaction and episode boundaries.}
Reset supplies the initial observation $o_0$. At decision $t=1,\ldots,H$, the environment decodes the action, advances the simulator by $\Delta t$ years, and returns $o_t$, reward $r_t$, execution status, and episode-end flags. An episode that reaches the configured endpoint has \texttt{truncated=True}. A reported solver failure sets \texttt{terminated=True} and returns a failure penalty; other execution failures are also recorded as failed attempts. Appendix~\ref{app:interface} specifies the penalty. Section~\ref{sec:views} describes the delayed-feedback variant of this interaction.

\subsection{Objectives and recorded outcomes}
\label{sec:objectives}

For successfully executed periods, the scalar reward combines climate, cost, equity, and electricity-price stability components:
\begin{equation}
 \begin{aligned}
 r_t={}&w_{\mathrm{clim}}r_t^{\mathrm{climate}}
       +w_{\mathrm{cost}}r_t^{\mathrm{cost}}\\
      &+w_{\mathrm{eq}}r_t^{\mathrm{equity}}
       +w_{\mathrm{stab}}r_t^{\mathrm{stability}},
 \qquad R=\sum_{t=1}^{H}r_t.
 \end{aligned}
 \label{eq:scalar-reward}
\end{equation}
The complete-episode return $R$ is undiscounted. The components are
\begin{equation}
 \begin{aligned}
 r_t^{\mathrm{climate}}&=-\max(0,T_t-T_{\mathrm{ref}})^2,\\
 r_t^{\mathrm{cost}}&=-\alpha_C C_t,
 \qquad r_t^{\mathrm{equity}}=-G_t,\\
 r_t^{\mathrm{stability}}&=-\alpha_E\max\!\left(0,
 \frac{E_t}{E_{t-1}}-1-\delta_E\right).
 \end{aligned}
 \label{eq:reward-components}
\end{equation}
Here $T_t$ is the recorded Hector temperature, $C_t$ and $G_t$ are the proxies defined below, and $E_t$ is the electricity-price index. The climate term penalizes temperatures above $T_{\mathrm{ref}}$; the stability term penalizes relative price increases above $\delta_E$. The coefficients $\alpha_C$ and $\alpha_E$ scale the component penalties before weighting. All three evaluation settings use the fixed parameters in Table~\ref{tab:interface-parameters}.

The reward threshold $T_{\mathrm{ref}}$ applies at each period; the distinct endpoint threshold $T_{\mathrm{goal}}$ defines target attainment. Both operate directly on Hector's emitted \texttt{global\_tas} channel, without an additional anomaly-baseline adjustment. The climate reference period for this archived configuration has not been independently established, so these thresholds define tests on the recorded channel. Appendix~\ref{app:interface} describes the temperature reader and missing-data behavior.

\begin{table}[t]
\centering
\caption{Fixed interface and objective parameters used in the reported evaluations. Price values follow the native monetary convention described in the text; temperature thresholds apply to the recorded Hector channel.}
\label{tab:interface-parameters}
\small
\begin{tabularx}{\linewidth}{lXl}
\toprule
Parameter & Meaning & Value\\
\midrule
$H$ & Decision periods per complete episode & $7$\\
$\Delta t$ & Duration of each period & $5$ years\\
$N$ & Regions in the emissions aggregation & $5$\\
$p_{\max}$ & API carbon-price ceiling & $1000$ per tCO$_2$\\
$\beta$ & Curvature of the action decoder & $3$\\
$\kappa$ & Carbon-mass price conversion factor & $44/12$\\
$(w_{\mathrm{clim}},w_{\mathrm{cost}},w_{\mathrm{eq}},w_{\mathrm{stab}})$ & Reward weights & $(0.40,0.30,0.15,0.15)$\\
$T_{\mathrm{ref}}$ & Period-wise climate penalty threshold & $1.5\degreeC$\\
$T_{\mathrm{goal}}$ & Endpoint target threshold & $2\degreeC$\\
$\alpha_C,\alpha_E$ & Cost and stability penalty scales & $10,\ 5$\\
$\delta_E$ & Relative price-increase threshold & $0.05$\\
\bottomrule
\end{tabularx}
\end{table}

\paragraph{Cost proxy.}
Let $Q_t$ denote summed native CO$_2$ query values and $Y_t$ summed \texttt{GDP\_Trial} demand. We use a clipped revenue-to-output proxy and its time accumulation:
\begin{equation}
 C_t=\operatorname{clip}_{[0,1]}
 \left(\frac{p_t\kappa\max(Q_t,0)}{Y_t}\right),
 \qquad K=\Delta t\sum_{t=1}^{H} C_t.
 \label{eq:cost-proxy}
\end{equation}
The numerator approximates carbon-tax revenue by multiplying the prescribed price, expressed per tonne of carbon, by nonnegative aggregate emissions. The denominator is the queried output measure. The implementation substitutes zero for the ratio when $Y_t$ is zero or NaN. We report $K$ in proxy-fraction years under this unit convention; it measures accumulated revenue relative to output, not discounted consumption loss or calibrated welfare. Native query conversions are specified in Appendix~\ref{app:interface}.

\paragraph{Distributional proxy.}
For each of the $N$ R5 regions, we clip aggregate emissions below at zero and divide by population after unit conversion to obtain per-capita emissions $z_{t,r}$. The regional mapping and conversion are specified in Appendix~\ref{app:interface}. Sorting these values in increasing order as $z_{t,(r)}$, we calculate the equal-region Gini coefficient
\begin{equation}
 G_t=\frac{\sum_{r=1}^{N}(2r-N-1)z_{t,(r)}}
 {N\sum_{r=1}^{N}z_{t,(r)}}.
 \label{eq:distribution-proxy}
\end{equation}
The coefficient is set to zero when the sum is nonpositive. It describes variation among regional per-capita emissions values with equal weight per region; it does not measure within-region inequality or the distribution of policy costs.

\paragraph{Attempt records.}
Each evaluation record is indexed by method, setting, seed, and attempt, and links the proposed policy to executed actions, feedback, and execution status. Failed attempts consume budget and retain their available trajectory prefixes. The reported incumbent is the best complete-episode return. For subsequent proposals, the available CMA-ES and BO adapter implementations assign penalized scores to failures. Appendix~\ref{app:methods} specifies these scores and their source provenance.

\subsection{Three evaluation settings}
\label{sec:views}

The three settings share the simulator configuration, horizon, and scalar objective. Each has a separately executed set of method--seed evaluations. The first two emphasize different outcome summaries under immediate feedback; the third changes feedback timing.

\textbf{Target and cost.} For each seed, we record the best complete return, whether any complete policy satisfies $T_H\leq T_{\mathrm{goal}}$, and the minimum $K$ among its target-reaching policies. Here $T_H$ is the endpoint temperature. The minimum-cost target-reaching policy may differ from the policy with the best return. Target-reaching cost is undefined for a seed with no qualifying policy.

\textbf{Cost--temperature trade-offs.} We retain $(K,T_H)$ for every complete evaluation to describe the outcomes visited by each scalar-reward system. Figure~\ref{fig:tradeoffs} displays the full recorded point sets.

\textbf{Delayed feedback.} A wrapper delays observations and rewards delivered to the agent by one decision step while preserving the simulator's period sequence. It returns the reset observation and zero reward during warmup. At episode end, all pending rewards are summed into the final feedback reward and the queues are cleared. Evaluation uses the underlying rewards to compute $R$ in Equation~\ref{eq:scalar-reward}. Schedule-search systems precommit their actions within each episode; PPO and LLM-ReAct choose actions as observations arrive.

\section{Experimental Protocol}
\label{sec:protocol}

\subsection{Seven configured systems}
Table~\ref{tab:config} summarizes the seven systems and their evaluation budgets. Random search, the fixed heuristic, CMA-ES~\citep{cma}, GP-BO~\citep{botorch}, and LLM+BO select complete schedules; PPO~\citep{ppo} and LLM-ReAct~\citep{react} select actions from stepwise observations. The comparison includes each system's policy representation, initialization, pretraining, and feedback rules. Appendix~\ref{app:methods} gives the fixed schedules, additional optimizer settings, failure scores, and configuration sources.

\begin{table}[t]
\centering
\caption{System configurations and attempted GCAM evaluations per seed and setting. Initial candidate evaluations and failed attempts count toward these budgets; mock pretraining and LLM inference are separate. The schedule length $H$ is defined in Table~\ref{tab:interface-parameters}.}
\label{tab:config}
\footnotesize
\begin{tabularx}{\linewidth}{lXr}
\toprule
System & Search or control configuration & Attempts\\
\midrule
Random & Seeded monotone random walk; no return-based adaptation & 24\\
Heuristic & Three fixed $H$-period schedules; order rotated by seed & 3\\
CMA-ES & $H$ normalized coordinates; population 6, initial $\sigma=0.2$ & 24\\
GP-BO & Three Sobol initial points; single-task GP, standardized outputs, and log expected improvement & 24\\
PPO & One mock-trained model per seed; 30,000 requested training steps; deterministic native control & 5\\
LLM-ReAct & Stepwise observations and recent action/reward/status history; constrained JSON action & 3\\
LLM+BO & Ten LLM-generated monotone initial schedules, then the same GP/log-EI search as GP-BO & 24\\
\bottomrule
\end{tabularx}
\end{table}

\paragraph{Search and initialization.}
For random search, the seed and attempt index determine the sampled price path. For the heuristic, the seed changes the order of the fixed schedules. CMA-ES searches the full normalized schedule domain~\citep{cma}. The two BO systems use the same GP/log-expected-improvement implementation after initialization~\citep{botorch,logei}, also over the full domain. Their initial designs differ in both source and size (Table~\ref{tab:config}), so the GP-BO--LLM+BO comparison combines these two choices. Failed evaluations remain in the budget; their use in subsequent proposal generation is specified in Appendix~\ref{app:methods}.

\paragraph{Pretrained and feedback-driven control.}
PPO uses one mock-trained model for each of the ten seeds~\citep{ppo}. Each model receives 22 observation scalars and emits four normalized controls, clipped to $[0,1]$ for native evaluation. Its parameters remain fixed across repeated evaluations, with inactive controls handled as specified in Section~\ref{sec:resource}. Training-step allocations in Table~\ref{tab:config} are requested counts; Appendix~\ref{app:methods} reports checkpoint evidence for actual training steps. LLM-ReAct receives the agent-facing observation, task identifier, step index, and latest three action/reward/status entries~\citep{react}. It returns a four-element JSON action with inactive coordinates fixed to zero. The experimental LLM configuration is \texttt{gpt-5.6-sol} with reasoning effort \texttt{high}; retry limits and the disabled fallback policy are specified in Appendix~\ref{app:methods}.

\subsection{Allocation and statistical readouts}

\paragraph{Native allocation.}
An evaluation cell is one system--setting--seed combination. Each system has ten cells in each of the three separately executed settings defined in Section~\ref{sec:views}, giving 210 cells and 3,210 attempted evaluations. The first two settings use immediate feedback with the same simulator and reward configuration; the third introduces the one-step delay. Random search, CMA-ES, GP-BO, and LLM+BO form the common-budget group with 24 attempts per cell. Table~\ref{tab:endpoints} additionally reports the heuristic, PPO, and LLM-ReAct at their respective allocations in Table~\ref{tab:config}. Budgets count evaluation attempts, including failures, rather than total computation. Episode timing omits some preempted work and optimizer overhead; mock pretraining and LLM inference are excluded from the attempt counts.

\paragraph{Endpoints and uncertainty.}
For a fixed system and setting, let $\mathcal{C}_s(b)$ contain the complete evaluations among seed $s$'s first $b$ attempts. With $R_{s,j}$ denoting the complete episode return of attempt $j$, the seed-level incumbent is
\begin{equation}
 I_s(b)=\max_{j\in\mathcal{C}_s(b)}R_{s,j}.
 \label{eq:protocol-incumbent}
\end{equation}
It is undefined when $\mathcal{C}_s(b)$ is empty. We report the IQM across the ten seed-level incumbents only when all ten are defined, and show the available-seed count separately. The IQM retains the middle 50\% of empirical mass with fractional weights at quartile boundaries~\citep{rliable}. Pointwise 95\% percentile intervals use 10,000 bootstrap resamples of the ten seed-level incumbents, recomputing the IQM in each resample. Repeated evaluations of a PPO model contribute to its seed-level incumbent, not additional bootstrap units. Appendix~\ref{app:endpoints} gives the resampling seed and pairwise comparison procedures.

\paragraph{Target attainment and cost.}
A seed attains the target if at least one complete policy satisfies $T_H\leq T_{\mathrm{goal}}$. We report the number of such seeds out of ten. For each target-reaching seed, the cost endpoint is the minimum accumulated proxy $K$ among its qualifying policies. Its all-seed IQM is reported only when every seed has at least one qualifying policy; otherwise the cost summary is undefined, even if all seeds have complete returns. The temperature threshold and cost proxy are defined in Section~\ref{sec:objectives}.

\paragraph{Delayed-feedback contrasts.}
At each system's allocated budget $b$, we match seed labels across the target-and-cost and delayed-feedback settings and compute
\begin{equation}
 \Delta_s=I_s^{\mathrm{delayed}}(b)-I_s^{\mathrm{target}}(b).
 \label{eq:protocol-delay}
\end{equation}
These post-hoc comparisons summarize the ten differences by their median, range, and sign counts (Appendix~\ref{app:endpoints}). They describe differences between separately executed runs. Matching labels does not control all variation in policy generation, simulator execution, or interruptions, so the contrasts do not identify a causal effect of delay.

\section{Evaluation Results}
\label{sec:results}


\begin{table}[t]
\centering
\caption{Evaluation results at the attempt budgets in Table~\ref{tab:config}. Best complete return is summarized by its IQM across ten seeds and a 95\% seed-bootstrap interval; higher is better. Completion is reported as completed/all attempts. Target counts are seeds with at least one complete policy satisfying $T_H\leq T_{\mathrm{goal}}$ (Section~\ref{sec:objectives}).}
\label{tab:endpoints}
\label{tab:complete}
\footnotesize
\begin{tabular}{lrrr}
\toprule
Method & Best return: IQM [95\% CI] & Completed / attempted & Target seeds \\
\midrule
\multicolumn{4}{l}{\textbf{Target and cost}} \\
Random & -0.918 [-0.945, -0.848] & 179/240 & 8/10 \\
Heuristic & -1.229 [-1.229, -1.229] & 10/30 & 0/10 \\
CMA-ES & -0.708 [-0.741, -0.667] & 194/240 & 10/10 \\
GP-BO & -0.900 [-0.948, -0.849] & 191/240 & 10/10 \\
PPO & -0.850 [-0.850, -0.850] & 50/50 & 0/10 \\
LLM-ReAct & -1.045 [-1.162, -0.978] & 24/30 & 6/10 \\
LLM+BO & -0.847 [-0.881, -0.818] & 178/240 & 10/10 \\
\midrule
\multicolumn{4}{l}{\textbf{Cost--temperature trade-offs}} \\
Random & -0.918 [-0.944, -0.848] & 179/240 & 8/10 \\
Heuristic & -1.229 [-1.229, -1.229] & 10/30 & 0/10 \\
CMA-ES & -0.724 [-0.759, -0.686] & 194/240 & 10/10 \\
GP-BO & -0.885 [-0.925, -0.830] & 190/240 & 10/10 \\
PPO & -0.850 [-0.850, -0.850] & 50/50 & 0/10 \\
LLM-ReAct & -1.032 [-1.171, -0.934] & 26/30 & 8/10 \\
LLM+BO & -0.821 [-0.828, -0.820] & 185/240 & 10/10 \\
\midrule
\multicolumn{4}{l}{\textbf{Delayed feedback}} \\
Random & -0.918 [-0.944, -0.848] & 179/240 & 8/10 \\
Heuristic & -1.229 [-1.229, -1.229] & 10/30 & 0/10 \\
CMA-ES & -0.708 [-0.741, -0.668] & 196/240 & 10/10 \\
GP-BO & -0.885 [-0.924, -0.830] & 190/240 & 10/10 \\
PPO & -0.850 [-0.850, -0.850] & 50/50 & 0/10 \\
LLM-ReAct & -1.021 [-1.100, -0.989] & 29/30 & 8/10 \\
LLM+BO & -0.839 [-0.873, -0.798] & 179/240 & 10/10 \\
\bottomrule
\end{tabular}

\end{table}

\subsection{Recorded returns and search progress}

CMA-ES has the highest best-return IQM among the four 24-attempt systems: $-0.708$ in target and cost, $-0.724$ in cost--temperature trade-offs, and $-0.708$ under delayed feedback (Table~\ref{tab:endpoints}). LLM+BO has a higher IQM than GP-BO in each setting. The archived rank-based tests do not detect a difference between their seed-level endpoints after Holm correction across 18 comparisons (adjusted $p=0.7099$ in each setting; Appendix~\ref{app:endpoints}). These tests address endpoint distributions rather than the IQM difference specifically.

Figure~\ref{fig:curves} in Appendix~\ref{app:endpoints} tracks best complete return against attempt budget. PPO's flat trace reflects repeated evaluation of fixed policies; the heuristic's all-seed IQM is defined only at its final budget. Figure~\ref{fig:allcurves} shows all three settings and return availability.

\subsection{Target cost and cost--temperature outcomes}

In the target-and-cost setting, CMA-ES, GP-BO, and LLM+BO each find a policy satisfying $T_H\leq T_{\mathrm{goal}}$ in all ten seeds. Their IQMs of the minimum qualifying cost proxy $K$ are respectively $0.333$, $0.458$, and $0.434$ proxy-fraction years, with 95\% intervals $[0.276,0.380]$, $[0.414,0.507]$, and $[0.373,0.479]$. Random search meets the criterion in eight seeds and LLM-ReAct in six; the heuristic and PPO meet it in none. Only the first three systems therefore have defined all-seed target-cost summaries under the protocol in Section~\ref{sec:protocol}.

The separately executed cost--temperature setting yields 834 complete coordinate pairs $(K,T_H)$ (Figure~\ref{fig:tradeoffs}). PPO's 50 pairs coincide at $(0,2.256\degreeC)$, and the heuristic's ten complete pairs coincide at $(0.442,2.415\degreeC)$; both temperatures exceed $T_{\mathrm{goal}}$. The search systems and LLM-ReAct produce more varied coordinates. Repeated outcomes remain in the record, with overlapping markers in the figure.

The pooled point set has 15 distinct nondominated coordinates: no other recorded point has both no greater cost and no higher temperature, with a strict improvement in at least one. PPO occupies the zero-cost end, while negative-temperature points from GP-BO and LLM+BO also enter the set. The runtime defects documented in Section~\ref{sec:integrity} can change these dominance relations.

\subsection{Completion and target attainment}

Among the 3,210 attempts, 2,493 are recorded as complete. Post-collection diagnosis classifies the remaining 717 as 714 numerical failures and three infrastructure failures; Appendix~\ref{app:record} maps these categories to the original worker statuses. Figure~\ref{fig:reliability} reports completion fractions per attempt and target-attainment fractions per seed at each system's allocated budget. PPO completes all evaluations without meeting the temperature criterion. In the target-and-cost setting, CMA-ES completes 194 of 240 attempts and finds a qualifying policy in every seed. Execution completion alone therefore does not guarantee target attainment.

\paragraph{Delayed feedback.}
The median seed-level contrast $\Delta_s$ is $-0.014$ for LLM-ReAct and $+0.016$ for LLM+BO; PPO retains zero carbon-price actions. Appendix~\ref{app:endpoints} gives the ranges, sign counts, and executed-sequence comparisons.

\subsection{Measurement Scope}
\label{sec:integrity}

The runtime diagnostics identified state-mutating residual checking, a residual-scale error, and recovery/read paths that bypass required validation. Omitting the checking path in a matched diagnostic changed one policy's recorded endpoint temperature from $-0.6625\degreeC$ to $1.964\degreeC$; this was not a validated replacement runtime (Appendix~\ref{app:record}). Such changes can affect rewards, target qualification, and nondominated outcomes. Their effect on the relative performance of the evaluated methods remains unresolved. A physically validated comparison requires new evaluations and adaptive search histories under a qualified runtime; rescoring selected endpoints alone is insufficient. Appendix~\ref{app:access} describes the replay resource and native-reuse requirements.

\section{Limitations}
\label{sec:limitations}

The experiments evaluate global carbon-price control through 2060 under a single GCAM configuration, fixed proxy objectives, and system-specific attempt budgets. Runtime measurement issues are documented in Section~\ref{sec:integrity}.

\section{Conclusion}

GCAM-Gym combines a shared policy-evaluation interface, configured search and control systems, and attempt-level records with numerical replay. In the archived runs, CMA-ES has the highest best-return IQM among the four equal-attempt-budget systems, and execution completion does not guarantee target attainment. The resource supports inspection of these recorded comparisons; validating policy outcomes and method rankings requires a qualified runtime and new evaluations with clean adaptive search histories.
\clearpage
\newpage
\section*{AI use statement}

We used generative AI tools to assist with coding, experiment management, manuscript editing, and draft figure illustrations. The authors reviewed the resulting code, experimental settings, text, and figures and take responsibility for the final manuscript and accompanying artifacts.

\clearpage
\appendix
\section{Space Bounds, Unit Conventions, and Query Fallbacks}
\label{app:interface}
\label{app:reward}

Table~\ref{tab:schema} gives the observation schema; Section~\ref{sec:objectives} defines the objective construction. This appendix specifies field dimensions and units, declared space bounds, additional unit conventions, and query fallbacks.

\begin{table}[t]
\centering
\caption{Observation schema. Sizes count scalar coordinates. Cost and distributional proxies are defined in Section~\ref{sec:objectives}; field ordering, unit conversions, and missing-data behavior are specified in Appendix~\ref{app:interface}.}
\label{tab:schema}
\footnotesize
\begin{tabularx}{\linewidth}{lrlX}
\toprule
Field & Size & Units & Construction\\
\midrule
\texttt{year} & 1 & calendar year & Simulator decision year\\
\texttt{temperature} & 1 & $\degreeC$ & Hector \texttt{global\_tas}, without rebasing\\
\texttt{co2\_concentration} & 1 & ppmv & Hector CO$_2$ channel\\
\texttt{emissions\_global} & 1 & GtCO$_2$/yr & Sum of five regional emissions\\
\texttt{emissions\_r5} & 5 & GtCO$_2$/yr & Mapped R5 aggregate emissions\\
\texttt{elec\_mix} & 7 & fractions & Electricity-generation shares\\
\texttt{energy\_price\_index} & 1 & ratio & Electricity price relative to reset reference\\
\texttt{policy\_cost\_frac\_gdp} & 1 & proxy fraction & Clipped revenue-to-output ratio $C_t$\\
\texttt{cdr\_deployed} & 1 & GtCO$_2$/yr & Aggregate sequestration, converted to CO$_2$ mass\\
\texttt{inequality\_gini} & 1 & dimensionless & Equal-region Gini $G_t$ of per-capita emissions\\
\texttt{carbon\_price\_effective} & 1 & price/tCO$_2$ & Native market price divided by $\kappa$\\
\texttt{solver\_iterations} & 1 & placeholder & Constant zero\\
\bottomrule
\end{tabularx}
\end{table}

\paragraph{Declared observation space.}
The environment declares a Gymnasium \texttt{Dict} of \texttt{Box} spaces with the componentwise bounds in Table~\ref{tab:obsbounds}. Year and solver iterations use integer coordinates; other fields use floating-point coordinates. These are declared API bounds; observations that violate them remain in the archive (Appendix~\ref{app:record}). Regional arrays follow the order \texttt{R5\_OECD90+EU}, \texttt{R5\_REF}, \texttt{R5\_ASIA}, \texttt{R5\_MAF}, and \texttt{R5\_LAM}. The seven electricity-share coordinates are coal, gas, oil, nuclear, hydro, wind/solar, and bioenergy/CCS/other. The worker source bundled with the resource specifies the mapping from native regions and subsectors to these groups. Flattening follows the Gymnasium dictionary-space order; LLM serialization sorts the field keys.

\begin{table}[h]
\centering
\caption{Declared observation bounds in the archived environment. Bounds apply separately to every coordinate of vector-valued fields; field dimensions and units are given in Table~\ref{tab:schema}.}
\label{tab:obsbounds}
\small
\begin{tabular}{lrr}
\toprule
Field & Lower bound & Upper bound\\
\midrule
\texttt{year} & 2025 & 2100\\
\texttt{temperature} & 0 & 10\\
\texttt{co2\_concentration} & 300 & 1500\\
\texttt{emissions\_global} & $-100$ & 150\\
\texttt{emissions\_r5} & $-50$ & 100\\
\texttt{elec\_mix} & 0 & 1\\
\texttt{energy\_price\_index} & 0.1 & 10\\
\texttt{policy\_cost\_frac\_gdp} & 0 & 1\\
\texttt{cdr\_deployed} & 0 & 25\\
\texttt{inequality\_gini} & 0 & 1\\
\texttt{carbon\_price\_effective} & 0 & 1000\\
\texttt{solver\_iterations} & 0 & $10^6$\\
\bottomrule
\end{tabular}
\end{table}

\paragraph{Action decoding and native units.}

Equation~\ref{eq:action-decoder} defines the decoder. For example, the active coordinate $a_t=0.5$ decodes to $p_t=182$. The API field \texttt{carbon\_price\_usd} is written to the native carbon market after multiplication by $\kappa$; the reported effective carbon price divides the queried market price by the same factor. The monetary convention and its validation limit are stated in Section~\ref{sec:resource}.

\paragraph{Query conversions.}
The worker converts native emissions and sequestration query sums to reported GtCO$_2$/yr by multiplying by $\kappa$ and dividing by $10^3$. For per-capita emissions, it multiplies the reported regional emissions by $10^9$ and the native population values by $10^3$ before division. These factors implement an MtC-to-GtCO$_2$ conversion and a thousands-of-people-to-people conversion. The cost proxy in Equation~\ref{eq:cost-proxy} instead uses the native emissions sum $Q_t$ and native \texttt{GDP\_Trial} demand sum $Y_t$ directly. Its interpretation as a revenue-to-output fraction requires compatible monetary units for the price and output queries; independent dataset-level validation remains outstanding.

\paragraph{Temperature channel.}
Hector is a reduced-complexity global climate model~\citep{hector}. The reader selects the last matching non-spinup \texttt{global\_tas} entry for the requested year from Hector's output stream. Section~\ref{sec:objectives} defines how this channel enters the reward and target criterion, including the unresolved reference-period convention. Re-execution requires the corresponding Hector configuration.

\paragraph{Electricity-price index and failure penalties.}
The electricity-price index divides the mean queried electricity-market price by the first available reference price, normally obtained at reset. A zero reference produces an index of one; a missing current price can produce NaN. The stability expression in Equation~\ref{eq:reward-components} has no additional guard for a zero previous index. The evaluation return is defined in Equation~\ref{eq:scalar-reward}; PPO's training discount is specified separately below.

On a reported solver failure, the environment replaces the usual reward components with scaled running component minima. Each minimum is initialized at $-10^{-3}$ and updated after successfully executed periods; the failure components are the running minima multiplied by $\lambda_{\mathrm{fail}}=1.5$. The scalar penalty uses the same reward weights as a successful period. These interaction penalties are distinct from the optimizer training scores described in Appendix~\ref{app:methods}.

\paragraph{Missingness and execution status.}
The archived worker also contains query fallbacks: absent regional emissions default to zero, missing regional population entries default to 1.0 before the per-capita unit conversion, unavailable sequestration may become zero, and unavailable effective carbon price may become zero. These are implementation behaviors, not independent observations of zero. In particular, the historical temperature reader could use a previous-year value when a requested observation was unavailable; this behavior is one of the defects in Appendix~\ref{app:record}. Environment failure penalties influence interaction, while incomplete trajectories have no complete-policy incumbent value in the reported analysis.

\section{Baseline Configuration and Feedback}
\label{app:methods}

Table~\ref{tab:config} summarizes the complete systems. The settings below supplement the main protocol with exact schedules, training parameters, and implementation provenance.

\paragraph{Schedule-search settings.}
The inspected adapter's random walk starts uniformly at an integer in $[0,40]$ and adds independent integer increments from $[0,45]$, capped at 1,000. The seeded generator depends on both the seed cell and evaluation index. CMA-ES~\citep{cma} has recorded bounds $[0,1]^7$, initial standard deviation 0.2, and population size six; the inspected adapter sets the initial center to $(0.25,\ldots,0.25)$. With 24 attempts, the allocation covers four populations. These coordinates are decoded through the shared nonlinear price map.

GP-BO has three recorded Sobol initial points. The inspected implementation uses BoTorch's \texttt{SingleTaskGP} with one-output standardization~\citep{botorch} and log expected improvement~\citep{logei}, limits fitting to 25 optimizer iterations, and selects from a 128-point scrambled Sobol candidate pool. LLM+BO uses ten initial candidates from the recorded LLM warm-start manifest, followed by the same GP/log-EI implementation. Its initial candidates are constrained to nondecreasing integer price schedules; the later BO search uses the full normalized coordinate box. Exact fitting settings are source-level descriptions, not all independently locked historical metadata.

In the inspected adapter, a failed BO observation at zero-based attempt index $j$ receives training utility $-10^6-j$; a complete observation receives its recorded return. CMA-ES minimizes negative complete return, with failed fitness $10^6+j$. These proposal-generation labels are not saved as checkpoint fields and differ from the environmental step penalty. The GP delegates default covariance and likelihood choices to its installed dependency; native re-execution must bind the adapter and dependencies. The replay shipped here uses saved records and performs no refitting.

\paragraph{Fixed schedules.}
The heuristic evaluates the following seven-period integer price sequences, rotating their execution order by seed:
\begin{align*}
\mathrm{mild}&=(30,47,68,91,116,150,190),\\
\mathrm{moderate}&=(45,60,80,105,130,165,205),\\
\mathrm{low}&=(0,1,3,5,7,12,19).
\end{align*}

\paragraph{PPO training and evaluation.}
The ten training manifests contain ten distinct model hashes. Each requests 30,000 mock-training steps with four environments, \texttt{MlpPolicy}, rollout length 256, batch size 64, ten epochs, learning rate $3\times10^{-4}$, and discount $\gamma=0.99$~\citep{ppo}. The inspected seed-zero checkpoint records 30,720 actual timesteps. The training environment has 22 observation scalars and four action coordinates. Native evaluation uses deterministic prediction and clips actions to $[0,1]$; unsupported coordinates can remain nonzero in the records but do not produce native interventions. The five evaluations per model and view are not additional training runs.

\paragraph{LLM control.}
The stepwise controller adapts the ReAct interaction pattern~\citep{react}. It receives the current agent-facing observation, step index, task identifier, and the most recent three action/reward/status entries. Its response schema contains only a four-element normalized \texttt{action}; unsupported coordinates must be zero. It does not emit a public interleaved reasoning transcript. The configured model is \texttt{gpt-5.6-sol}, reasoning effort \texttt{high}, with a 2,048-token response limit, 60-second timeout, at most two retries, and a six-second minimum interval. Fallback policies are disabled. Deployment endpoints and credentials are excluded from the resource files.

\paragraph{Configuration provenance.}
Per-attempt metadata and frozen training and warm-start manifests establish recorded settings. Other details, including the CMA initial center, GP fitting settings, and optimizer failure labels, describe the inspected adapter but lack a contemporaneous binding of every implementation byte. The archived package preserves environment and worker definitions. The selected export omits native binaries and optimizer dependencies; the README distinguishes these provenance levels from portable arithmetic replay.

\section{Statistical Details and Additional Views}
\label{app:endpoints}

Table~\ref{tab:endpoints} reports results for all evaluated methods with endpoint intervals, completion counts, and target attainment. Here we give the statistical procedures, three-setting evaluation curves, descriptive delayed-feedback contrasts, and action-record details.

\begin{figure}[t]
\centering
\includegraphics[width=\linewidth]{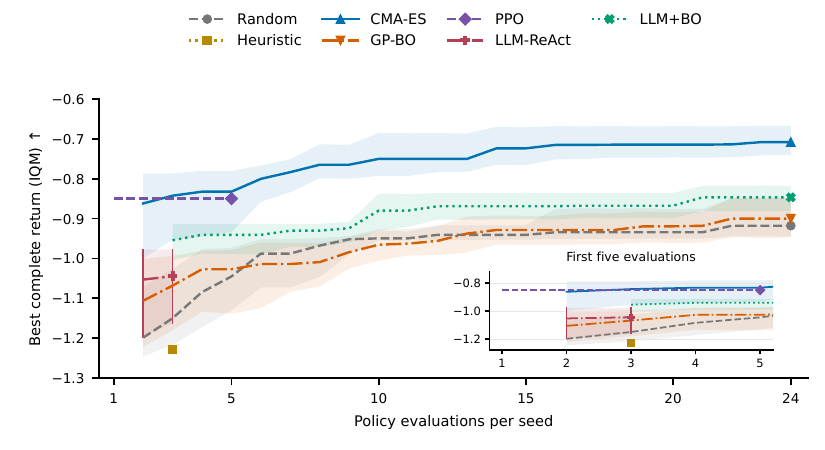}
\caption{Best complete return versus attempted evaluations in the target-and-cost setting. Curves show ten-seed IQMs and pointwise 95\% seed-bootstrap intervals; short traces retain vertical interval bars. Markers identify each system's final budget, and the inset enlarges the first five attempts. Failed attempts count toward the budget, and estimates remain unplotted until all ten seeds have a complete return. The axis excludes pretraining and LLM inference; PPO parameters remain fixed.}
\label{fig:curves}
\end{figure}

All 21 endpoint rows are retained. At prefixes where any seed has no complete policy, the all-seed return IQM is undefined. Minimum target-cost IQMs similarly require all ten seeds to have a target-reaching policy. The native record's numerical failure taxonomy and the measurement boundary apply to these saved outcomes. Degenerate observed bootstrap intervals do not imply certainty about another training run or another simulator configuration.

For the heuristic, complete returns are available for four, seven, and ten seeds after the first, second, and third attempts, respectively, in each setting. Its all-seed IQM is therefore defined only at the final budget. The seed-dependent ordering of the fixed schedules produces this availability pattern.

The archived analysis uses 10,000 percentile seed-bootstrap resamples with RNG seed 20260916. Its IQM retains the middle 50\% empirical mass with fractional boundary weights~\citep{rliable}. The protocol was fixed after collection and before the first comparative analysis. Six pairwise search comparisons per view form an 18-comparison family with Holm adjustment~\citep{holm} for exact two-sided pooled-average-rank label-permutation tests. A separate family uses paired seed-label sign flips as a dependence sensitivity. GP-BO versus LLM+BO has adjusted rank-test $p=0.7099$ in each view. These saved intervals and tests are retained; the portable replay independently reconstructs their input summaries, not the resampling or permutation procedures.

For each rank test, the statistic is the absolute deviation of the first group's rank sum from $n(N+1)/2$, with average ranks for ties, $n=10$, and $N=20$. All $\binom{20}{10}$ label allocations are enumerated. The paired sensitivity instead uses the absolute mean of ten seed-label differences and enumerates all $2^{10}$ sign assignments. Comparisons include numerical tolerances of $10^{-10}$ and $10^{-12}$, respectively. Within each 18-test family, sorted raw probabilities $p_{(j)}$ receive adjusted values $\widetilde p_{(j)}=\min\{1,\max_{i\leq j}(19-i)p_{(i)}\}$~\citep{holm}.

\begin{figure}[t]
\centering
\includegraphics[width=\linewidth]{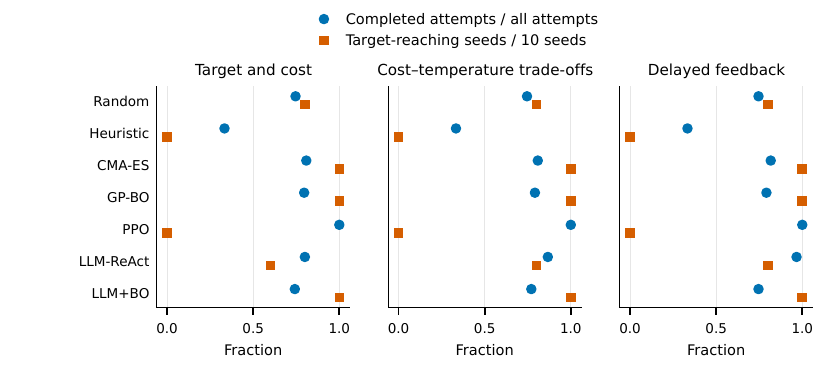}
\caption{Completion and target attainment across the evaluated methods. Circles denote completed attempts divided by all attempts; squares denote seeds with any complete policy satisfying $T_{2060}\leq2\degreeC$, divided by ten. Shapes are vertically offset to preserve coincident observations. The native budgets are those in Table~\ref{tab:endpoints}; proportions describe this recorded sample.}
\label{fig:reliability}
\end{figure}

\begin{figure}[t]
\centering
\includegraphics[width=\linewidth]{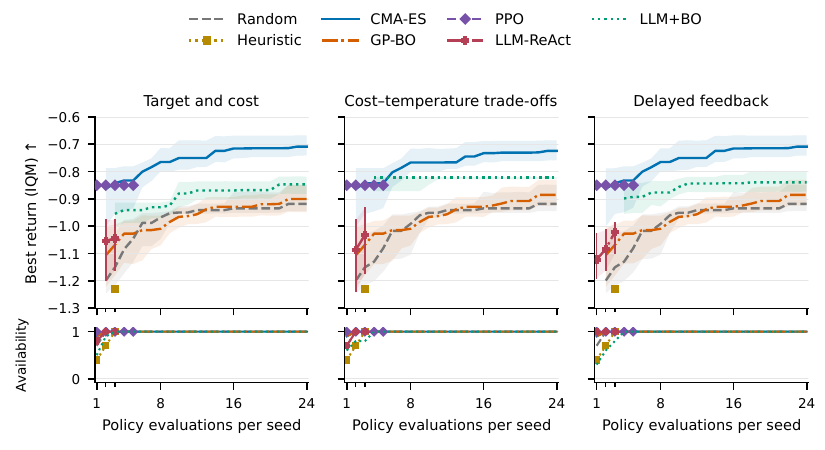}
\caption{Evaluation progress for all evaluated methods across the three settings. Upper panels show the IQM of best complete return across ten seeds with saved pointwise 95\% seed-bootstrap intervals; short traces also show vertical intervals. Lower panels show the fraction of seeds with a complete policy. The horizontal axes count policy evaluations per seed, including failures. Traces stop at each system's actual budget, and undefined all-seed estimates remain unplotted. PPO traces describe evaluation of mock-pretrained policies, not native learning; pretraining and LLM inference costs are excluded.}
\label{fig:allcurves}
\end{figure}

\begin{table}[t]
\centering
\caption{Post-hoc return contrasts between the delayed-feedback and target-and-cost settings at each system's allocated budget, using ten matched seed labels per system. Positive $\Delta$ denotes higher best complete return under delayed feedback. Values are displayed to three decimals; sign counts use unrounded differences with a $10^{-12}$ tolerance for floating-point roundoff, not an equivalence test. The contrasts do not isolate a causal delay effect.}
\label{tab:delay}
\small
\begin{tabular}{lrrrl}
\toprule
System & Median $\Delta$ & Min $\Delta$ & Max $\Delta$ & Higher / tied / lower \\
\midrule
Random & 0.000 & 0.000 & 0.000 & 0 / 10 / 0 \\
Heuristic & 0.000 & 0.000 & 0.000 & 0 / 10 / 0 \\
CMA-ES & 0.000 & 0.000 & 0.000 & 0 / 10 / 0 \\
GP-BO & 0.000 & 0.000 & +0.099 & 2 / 8 / 0 \\
PPO & 0.000 & 0.000 & 0.000 & 0 / 10 / 0 \\
LLM-ReAct & -0.014 & -0.185 & +0.228 & 4 / 0 / 6 \\
LLM+BO & +0.016 & -0.157 & +0.155 & 6 / 1 / 3 \\
\bottomrule
\end{tabular}

\end{table}

\paragraph{Descriptive delayed-feedback contrasts.}
The median paired-label return difference $\Delta_s$ is zero for random search, the heuristic, CMA-ES, GP-BO, and PPO (Table~\ref{tab:delay}). LLM-ReAct has median $-0.014$, with four positive and six negative differences, spanning $[-0.185,0.228]$. Its target-reaching count nevertheless increases from six to eight seeds, illustrating that median return change and target discovery answer different questions. LLM+BO has median $+0.016$, spanning $[-0.157,0.155]$. Its schedule search does not use stepwise feedback, so the latter contrast is not evidence of a within-episode delay response.

\paragraph{Delay timing and action records.}
For a seven-step delayed-feedback episode, the agent receives observation sources $(0,1,2,3,4,5,6)$, where zero denotes the reset state. Rewards from steps one to five arrive one step late; the terminal call returns the queued rewards from steps six and seven. The task metrics use the actual trajectory record. Pairing same-numbered seed cells across the target-and-cost and delayed-feedback settings gives 1/0/2/2/1/10/10 cells with different executed four-coordinate arrays for Random/Heuristic/CMA-ES/GP-BO/PPO/LLM-ReAct/LLM+BO. Comparing only the active carbon-price coordinate gives 1/0/2/2/0/10/10; comparison of decoded price sequences gives the same counts. A shorter failed prefix also constitutes a different sequence, so these counts do not measure deliberate changes to full proposed schedules.

The action records distinguish changes in effective control from changes in inactive API coordinates. For LLM-ReAct, every matched seed cell contains at least one same-indexed attempt with a different executed carbon-price sequence. PPO emits a zero carbon-price coordinate in all 150 native evaluations: its single matched cell with a changed four-coordinate action array differs only in inactive controls. Thus, the PPO trace difference does not indicate a changed native carbon-price intervention. Differences in executed sequence length can also arise from failed prefixes; these paired-label comparisons describe the recorded behavior rather than isolate a causal effect of delay.

\paragraph{Cost--temperature point sets.}
The 834 points comprise 179/10/194/190/50/26/185 complete evaluations in method order. Every successful attempt in the trade-off setting contributes its recorded cost proxy and temperature; none is selected by its score or Pareto membership. Failed attempts contribute no complete coordinate pair. Repeated policies remain recorded separately, although their markers can overlap. The exported record preserves all successful and failed attempts, enabling alternative inspection without pretending that duplicate points are independent physical-policy observations.

\section{Runtime Evidence and Status Semantics}
\label{app:record}

The original worker statuses are 2,493 \texttt{success}, 716 \texttt{solver\_failure}, and one \texttt{api\_failure}. The post-collection diagnostic taxonomy reclassifies known execution exceptions, giving 2,493 successes, 714 numerical failures, and three infrastructure failures. These are two labelings of the same 3,210 attempts. The main tables retain the full grid. The recorded infrastructure sensitivity excludes three entire cells: CMA-ES seed 9 in the trade-off setting, CMA-ES seed 3 under delayed feedback, and LLM-ReAct seed 0 in the trade-off setting; it does not replace the main endpoint summaries.

The runtime audit first checked all 2,493 recorded successful trajectories. It found 57 trajectories outside the declared observation space, affecting 43 method--view--seed cells. These were interface-contract violations, not newly introduced physical rejection thresholds. The audited data were preserved without clipping temperatures or deleting those records.

In a matched native diagnostic, two fixed policies were each executed with the archived residual-checking path and with a diagnostic omission of that path. The policy, restart, configuration, native library, and single-thread execution convention were held fixed. For the zero-price policy, recorded 2060 temperature changed from $2.256\degreeC$ to $2.291\degreeC$. For the selected GP-BO policy, it changed from $-0.6625\degreeC$ to $1.964\degreeC$, with CO$_2$ changing from 113.7 to 467.5 ppm. The native CSV contained the anomaly, and the preserved worker completed without recovery, so the example cannot be explained solely as a JSON-reading or replay error. The omission mode was diagnostic and did not establish a replacement convergence test.

The audit additionally identified an extra quantity-scale factor in the old residual ratio and fail-open recovery/read paths, including substitution of previous-year climate observations. Subsequent repairs were versioned separately. A clean adaptive search under a repaired measurement process cannot be reconstructed by replacing only an old winner's return: earlier feedback also determines later proposals. The source-linked diagnostic summary shipped with the manuscript identifies the preserved records and the scope of these observations.

The selected-policy diagnostic changes the endpoint climate penalty because the two readings straddle $T_{\mathrm{ref}}$, although both satisfy $T_{\mathrm{goal}}$. More generally, affected measurements can alter rewards, target qualification, and the policies included in minimum-target-cost and nondominance summaries. Their effect on the relative performance of the evaluated methods has not been established. A physically validated comparison requires a qualified runtime and new evaluations. For adaptive search, it also requires rerunning proposal generation with the validated feedback: rescoring selected policies cannot recover the search histories that different earlier measurements would have produced.

\section{Reuse requirements}
\label{app:access}

Replay verifies the arithmetic from the selected records; saved bootstrap intervals and permutation tests are retained without reestimation. It does not rerun the native simulator. The archive omits native binaries, initialization artifacts, and trained PPO checkpoints; native re-execution requires those inputs, their dependencies, and applicable access permissions. Runtime and measurement validation are prerequisites for interpreting a new comparison physically. Broader controls and horizons beyond the present carbon-price study through 2060 would additionally require extending the interface and evaluation design.

\end{document}